\documentclass[aps,pra,reprint,superscriptaddress]{revtex4-1}

\usepackage{amsmath}  
\usepackage{amsfonts} 
\usepackage{graphicx,color,calc,adjustbox}
\usepackage[version=4]{mhchem}
\usepackage{natbib}
\definecolor{orange}{rgb}{1,0.5,0}

\newcommand{\cwidth}{\columnwidth}

\begin{document}

\title{Light diffraction from a phase grating at oblique incidence in the intermediate diffraction regime}

\author{Georg Heuberger} 
\email{Present address: Sperlgymnasium, 1020 Wien, Austria}
\affiliation{Faculty of Physics, University of Vienna, 1090 Wien, Austria}

\author{J\"urgen Klepp}
\email{Corresponding author: juergen.klepp@univie.ac.at}
\affiliation{Faculty of Physics, University of Vienna, 1090 Wien, Austria}

\author{Jinxin Guo}
\email{Present address: Institute of Information Photonics Technology, Faculty of Science, Beijing University of Technology, Beijing 100124, China}
\affiliation{Department of Engineering Science, University of Electro-Communications, 1-5-1 Chofugaoka, Chofu, Tokyo 182, Japan}

\author{Yasuo Tomita}
\affiliation{Department of Engineering Science, University of Electro-Communications, 1-5-1 Chofugaoka, Chofu, Tokyo 182, Japan}

\author{Martin Fally}
\affiliation{Faculty of Physics, University of Vienna, 1090 Wien, Austria}

\date{\today}

\date{\today}

\begin{abstract}
We experimentally characterize the positions of the diffraction maxima of a phase grating on a screen, for laser light at oblique incidence (so-called off-plane diffraction or conical diffraction). We discuss the general case of off-plane diffraction geometries and derive basic equations for the positions of the diffraction maxima, in particular for their angular dependence. In contrast to previously reported work [Jetty \emph{et al.}, Am.\,J.\,Phys. {\bf 80}, 972 (2012)], our reasoning is solely based on energy- and momentum conservation. We find good agreement of our theoretical prediction with the experiment.
A detailed discussion of the diffraction maxima positions, the number of diffraction orders, and the diffraction efficiencies is provided. We assess the feasibility of an experimental test of the phenomenon for neutron matter waves.
\end{abstract}

\maketitle

\section{Introduction}
A standard approach to diffraction phenomena is based on the following two simple cases: First, diffracting light from optically thin gratings at normal incidence is considered. The numerous observable diffraction maxima that exhibit little dependence on the angle of incidence are usually explained in terms of multi-wave interference with the diffraction angles being governed by the grating equation (see, for instance, \cite{Halliday-07}). 
Second, Bragg diffraction from thick gratings is introduced within the context of determining crystal structures. In this case, normal incidence does not lead to any diffraction. Instead, the condition under which constructive interference occurs and a sharp diffraction maximum can be observed is given by Bragg's law \cite{Halliday-07}. 

\begin{figure*}
 \includegraphics[width=2\cwidth]{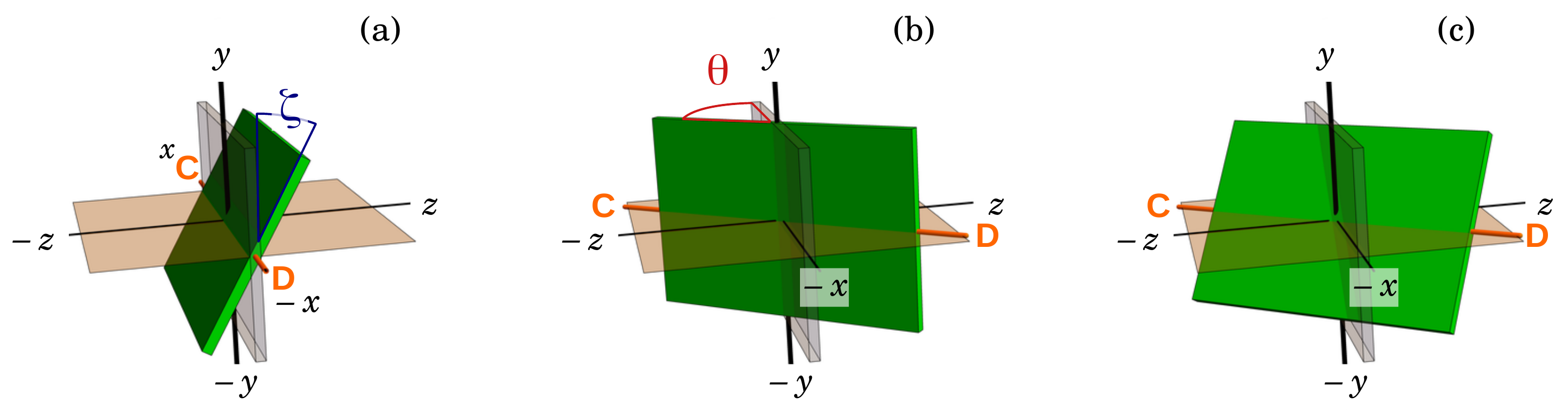}
\caption{\label{fig:coordsystem} The coordinate system $xyz$ is fixed to the lab frame such that the light wave is incident along the positive $z$-direction. To set oblique incidence, the grating (gray and green slabs) is tilted around the grating vector $\vec G$ (direction made visible by axis \textcolor{orange}{\sf CD}; fixed to the sample) by an angle $\zeta$ (a). To measure the angular dependence for a particular angle $\zeta$, the grating undergoes step-wise rotation through angles $\theta$ around the $y$-axis (b). A general situation is shown in (c). The values for $\zeta$ and $\theta$ are: (a) $\zeta=30^\circ$, $\theta=0^\circ$, (b) $\zeta=0^\circ$, $\theta=50^\circ$, (c) $\zeta=30^\circ, \theta=50^\circ$.} 
\end{figure*}

It is clear that the above two cases are two rather simple extremes of more general, complicated situations in diffraction physics. Theories for almost any conceivable configuration other than the two mentioned above are treated in the literature (see, e.g., \cite{Loewen-97,Palmer-14}), but remain widely unknown to most non-specialists. Here, we elucidate one of these general cases experimentally: oblique incidence on a (holographic) phase grating exhibiting diffraction in the so-called intermediate diffraction regime \cite{Gaylord-ao81} that is in between the Raman-Nath regime for optically thin gratings \cite{Raman-piasa36, Raman.2-piasa36} and the Bragg regime for optically thick gratings. We deploy a planar, one-dimensional unslanted grating whose diffraction properties cannot be described by either of the two extreme cases outlined above. 
First, we describe measurements of the angular dependence of the diffracted intensities for the simple and usual case of in-plane diffraction, i.e., for incoming and outgoing beams lying in the same plane. 
To vary the angle of incidence, the grating is rotated in steps by angles $\theta$ about an axis perpendicular to the latter plane (see Fig.\,\ref{fig:coordsystem}). As the next step, starting again from normal incidence, we tilt the grating around its grating vector by an angle $\zeta$ to obtain oblique incidence, and measure the angular dendence also for this more complicated situation.
The concept of employing extreme oblique incidence (`off-plane mount') is of importance not only for neutron optics \cite{Klepp-Materials12,Tomita-jmo16} but also 
for X-rays \cite{Seely-ao06} in designing spectrometers for space applications \cite{McEntaffer-sp08}, for instance, as well as for extreme UV light at grazing incidence \cite{Poletto-ao06,Goray-josaa10}.

Two particular questions are investigated in the present work: How many diffraction orders occur and where are the diffraction maxima located upon rotation or tilt?
In previous studies the above questions have already been answered to a certain extent \cite{Phadke-ajp86,Jetty-ajp12}: They find -- both thoretically and experimentally -- the positions of the diffraction maxima as a function of angles measured with respect to well-defined rotation axes, which are fixed to the diffraction grating. However, their theoretical approach is independent of the grating spacing and the incident wavelength used: The intensities of the diffracted beams are estimated by using the Fresnel-Kirchhoff formalism. In the far-field limit (Fraunhofer diffraction) the amplitudes of the diffracted beams are, therefore, proportional to the Fourier transform of the aperture function of the grating. The approach thereby implemented is equivalent to applying the first Born approximation \cite{Born-02}, valid when the refractive index of a medium does not differ too much from unity, negelecting any multiple diffraction processes. 
As a natural extension of the previous studies, in the present work we give a simple analytic expression for the location of the diffraction maxima upon rotation and tilt of a phase grating based on the Floquet condition \cite{Gaylord-apb82}, i.e. based on energy and momentum conservation only. Moreover, concerning the intensities of the diffracted beams and their angular dependence, we adopt a more general approach: While the Fresnel-Kirchhoff theory is well justified for the situation studied in Ref.\,\cite{Jetty-ajp12} (grating spacing $\Lambda\approx 83\,\mu$m, incident wavelength $\lambda=532$\,nm, by which diffraction is clearly restricted to the Raman-Nath regime \cite{Raman-piasa36,Raman.2-piasa36}), in our work we employ the rigorous coupled wave analysis (RCWA,\,\cite{Moharam-josa81,Gaylord-apb82,Moharam-josaa95a}) to solve the diffraction problem for a wavelength/grating/geometry combination, in which angular dependencies of the diffracted intensities can neither be treated in the Raman-Nath regime nor by the theory for thick gratings (Bragg regime). Finally, we also discuss the feasibility of a similar measurement by neutron matter waves.

\section{Modelling}

\begin{figure}
\centering
 \includegraphics[width=0.99\columnwidth]{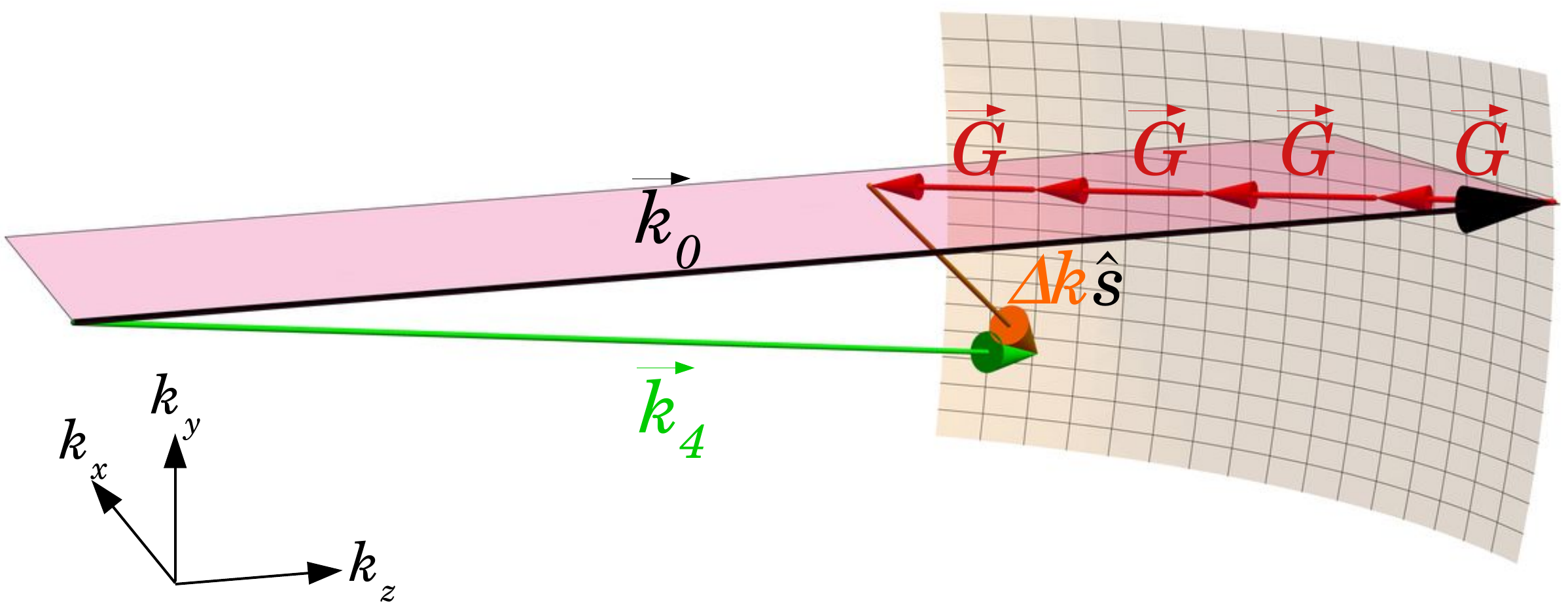}
\caption{\label{fig:wavevec} Wavevector diagram for the 4th order diffraction process in the oblique case with $\theta > 0^\circ$ and $\zeta > 0^\circ$ as in Fig.\,\ref{fig:coordsystem}\,(c).  
The curved surface is part of the Ewald-sphere. Note that $\vert\vec k_0\vert=\vert\vec k_4\vert$.}
\end{figure}

We shall consider a one-dimensional phase grating with the spatially modulated refractive index n(x) given by
\begin{equation}\label{eq:RefrIndProf}
n(\vec x)=n_0+n_1\cos(\vec G\cdot\vec x)+
n_2\cos(2\vec G\cdot\vec x)+\ldots,
\end{equation}
where $\vec G$ is the grating vector, $n_0$ is the average refractive index of the grating and the $n_{1,2,\dots}$ are the amplitudes of the various Fourier components, boils down to solving the associated boundary-value problem. Exact solutions were given in terms of a modal theory (often called dynamic diffraction theory, see, e.g., Ref.\,\cite{Russell-prep81}) or, alternatively, a coupled-wave theory \cite{Moharam-josa81}. The strategy is to solve Maxwell's equations in each of the regions (input, grating, output) and match the tangential components of the fields at the boundaries. The Floquet theorem, which singles out the permitted fields in a periodic lattice, requires $\vec q_m=\vec q_0-m\vec G$, where $\vec q_m$ and $\vec q_0$ are the wavevectors of the $m$-th diffraction order in the grating with $\vert\vec q_m\vert=2\pi n_0/\lambda$ and of the incident wave, respectively. In fact, for the spatially bounded grating with periodicity only along the $x$-direction (see Fig.\,\ref{fig:coordsystem}), it is sufficient that the wavevector component parallel to the sample surface and along the grating vector obeys Floquet's condition \cite{Gaylord-apb82}. Furthermore, at the boundaries this parallel component $\vec q_{||}$  matches the parallel component in the bounding medium (here: free space) $\vec k_{||}$, so that the permitted wavevectors of diffraction outside the grating are given by:
\begin{eqnarray}\label{eq:0}
\vec k_m&=&\vec k_0-m\vec G-\Delta k\hat s\\
|\vec k_m|&=&2\pi/\lambda\rightarrow\beta,\label{eq:1}
\end{eqnarray}
where $\hat s$ denotes the unit vector of the sample surface normal and $\Delta k$ is the off-Bragg dephasing parameter \cite{Sheridan-jmo92}, which we derive below. An example for oblique-incidence diffraction according to Eq.\,(\ref{eq:0}), illustrating the meaning of the involved physical quantities, is depicted in Fig.\,\ref{fig:wavevec}.

Equation\,(\ref{eq:0}) represents the diffraction condition or Laue equation (see, for instance, \cite{Ashcroft-76}). 
 The latter is equivalent to the aforementioned Bragg's law given by $2\Lambda\sin\theta_{m}=m\lambda$. 

Next, we provide analytic expressions for the diffracted wavevectors for geometry in our experiments. As sketched in Fig.\,\ref{fig:coordsystem}\,(a), a phase grating is placed on a sample holder which allows for tilting it around its grating vector $\vec G$ (collinear to axis CD) by a tilt angle $\zeta$. The sample holder is fixed on a rotation stage used to implement the rotation of the grating through angles $\theta$ about the rotation axis $y$ [see Fig.\,\ref{fig:coordsystem}\,(b)]. The $y$-axis is perpendicular to the grating vector and is fixed in the lab frame of reference, independent of the tilt angle $\zeta$ \footnote{In contrast to the experimental scenario discussed in the work of Jetty\,{\it et al.}\,\cite{Jetty-ajp12}, results of the present study are independent of the sequence of rotations. The present coordinate system relates to that of Ref.\,\cite{Jetty-ajp12} 
as $x\to-Y,y\to X,z\to Z$, the angle $\zeta\to\theta$. There is no equivalent for 
$\theta$ of Jetty's work. Our sequence of rotations ($\zeta$ followed by $\theta$) is equivalent to the one described by Fig.\,7\,(a) and Eq.\,(17) of Ref.\,\cite{Goray-josaa10}.}.
Also, here, the vectors $\vec k_0$, $\vec G(\theta)$, and $\hat s(\theta,\zeta)$ are given in the lab frame. Without loss of generality, we may choose our coordinate system such that the incoming beam corresponds to the wavevector $\vec k_0=(0,0,\beta)$. The grating vector remains in the $x-z$-plane and can be written as $\vec G=G(\cos\theta,0,-\sin\theta)$ [cf. Fig.\,\ref{fig:coordsystem}\,(b)], with $G=|\vec G|=2\pi/\Lambda$. Consulting Figs.\,\ref{fig:coordsystem}\,(b) and \ref{fig:coordsystem}\,(c), one can see that the grating surface normal may be given by $\hat s=(-\cos\zeta\sin\theta,\sin\zeta,-\cos\zeta\cos\theta)$. Thus, by combining Eqs.\,(\ref{eq:0}) and (\ref{eq:1}), $\Delta k$ can be expressed in terms of the angles $\theta$ and $\zeta$ as
 
\begin{eqnarray}
\Delta k\!&=&\!-\beta\!\left[\cos\theta\cos\zeta\!-\!\frac{1}{2}\right.\nonumber\\
 \!\!\!\!&\times&\left.\!\!\!\sqrt{
1\!\!+\!2\!\cos(2\zeta)\!\cos^2\!\theta\!+\!\cos(2\theta)\!-\!8\rho_m\!\sin\theta\!-\!4\rho_m^2
}\,\right]\!\!,\label{eq:mismatch}
\end{eqnarray}
where $\rho_m:=m G/\beta$.
Consequently, Eqs.\,(\ref{eq:0}) and (\ref{eq:mismatch}) completely determine the directions $\vec k_m/\beta$ of the diffracted beams for the $m$-th diffraction order. Note that this result has been derived only from momentum and energy conservation ($\vert\vec k_0\vert=\vert\vec k_m\vert$). It is expected from Eq.\,(\ref{eq:0}) that, for increasing $\Delta k$ the diffraction angles rapidly deviate from $2\theta_{m}$. Note that Eq.\,(\ref{eq:mismatch}) yields zero for $m=0$, meaning that the forward-diffracted beam is not expected to experience any deviation from the plane of incidence.

\section{Experiments}\label{sec:Experiments}

\subsection{Preparation of a tranmsission phase grating}
A holographic phase grating for the diffraction experiments was prepared by using a photopolymerizable nanoparticle composite material used for holographic applications \cite{Tomita-jmo16}. SiO$_2$ nanoparticles with an average diameter of 13\,nm and bulk refractive index of 1.46, dispersed in a solution of methyl isobutyl ketone, were mixed with methacrylate monomers (2-methyl-acrylic acid 2-4-[2-(2-methyl-acryloyloxy)-ethylsulfanylmethyl]-benzylsulfanyl-ethyl ester) \cite{Suzuki-ao04}. The refractive index of the formed polymer was 1.59 at 589\,nm. The doping 
concentration of SiO$_2$ nanoparticles was 34\,vol.\%. Photoinitiator titanocene (Irgacure 784, Ciba) was mixed at
1\,wt.\% with respect to the monomer to provide photosensitivity in the green. The chemical mixture
was cast on a glass plate, dried and covered with another glass plate, the latter separated
from the former by spacers of known thickness. A two-beam interference setup with two mutually coherent s-polarized beams of equal intensities from a laser
diode-pumped frequency-doubled Nd:YVO$_4$ laser oper-ating at 532\,nm was used to record a hologram providing an unslanted transmission phase grating with $\Lambda=5 \,\mu$m and the grating thickness $d\approx 13\,\mu$m: In the bright regions of the interference
pattern irradiating the sample, the photoinitiator triggers the polymerization process. Monomer is consumed in the bright regions by the polymer formation. As a consequence of the resulting chemical potential difference between the bright and dark regions, the mutual diffusion of monomer in the dark regions and nanoparticles in the bright regions finally results in the increased concentrations of nanoparticles in the dark regions and the formed polymer in the bright regions \cite{Tomita-jmo16}. Such a difference in their concentrations provides a spatially periodic modulation of the refractive index, i.e., a holographic phase grating.

\subsection{In-plane diffraction ($\mathbf{\zeta = 0^\circ}$)\label{sec:DE}}
\begin{figure}
\begin{center}
 \includegraphics[width=0.9\cwidth]{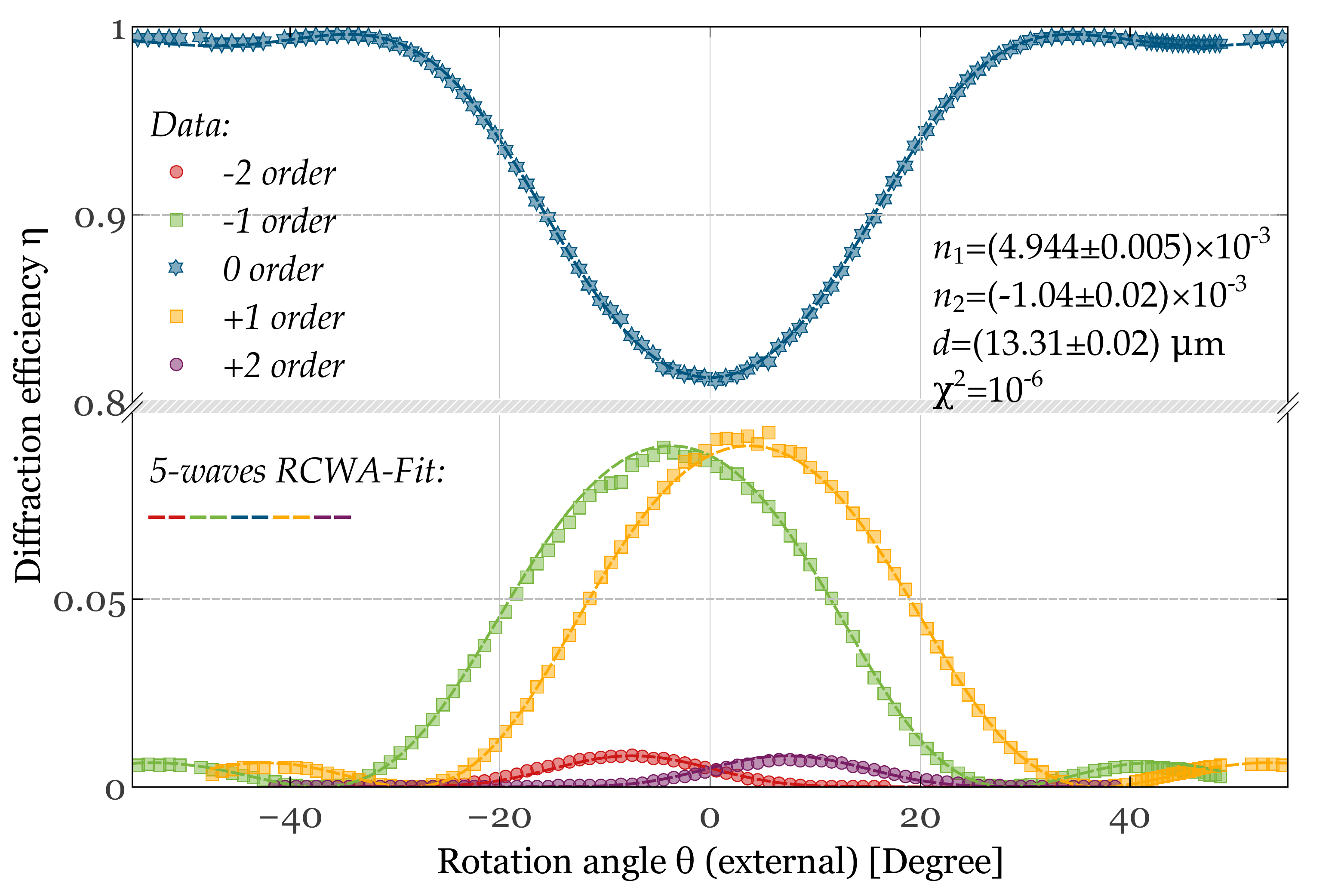}
\end{center}
\caption{\label{fig:DE} Angular dependence of the diffraction efficiency for the $\pm 2,\pm 1, 0$th orders at $\zeta=0^\circ$. Error bars for the experimental data are included but are much smaller than the symbols. Dashed lines are fits to the data using the RCWA (see text).} 
\end{figure}

The angular dependence of the diffracted intensities $I_m(\theta)$ of the $-3\ldots+3$ diffraction orders at $\zeta=0^\circ$ were measured by placing Si-photodiodes at the positions of the diffracted beams. For a pure phase grating the diffraction efficiency for diffraction order $m$ can be obtained from the measured data using the formula $\eta_m=I_m/\sum I_m$.
A plot of the diffraction efficiencies probed by a He-Ne laser (633 nm) is shown in Fig.\,\ref{fig:DE}. It can be clearly seen that diffraction for our grating cannot be described properly by the Raman-Nath theory, for the diffraction process exhibits considerable angular selectivity, i.e., the diffraction efficiency decreases substantially for $\theta$ not too far from the Bragg angle.
Neglecting rather lower $\pm$3rd-order signals, RCWA fits to the other order signals are shown by curves in Fig.\,\ref{fig:DE}. In particular, an approximate RCWA calculation was performed for only 5 diffraction orders of the pure phase grating.
Fit parameter estimations were found to be
$n_1=(4.944\pm0.005)\times 10^{-3}$, $n_2=(-1.04\pm 0.02)\times 10^{-3}$, and $d=(13.31\pm 0.02)\,\mu$m at $\chi^2=10^{-6}$, respectively. Considering the magnitudes of $n_1$ and
$n_2$, we consider that the refractive index profile
of the grating is not completely sinusoidal, as $|n_2|$ is
of the same order of magnitude as $|n_1|$. The minus sign of $n_2$ 
indicates that there is a phase shift of $\pi$ between the first
and the second Fourier components [cf. Eq.\,(\ref{eq:RefrIndProf})] of the refractive index profile of this grating.
The RCWA fitting is found to be in good agreement with the data. 

\subsection{Off-plane diffraction ($\mathbf{\zeta \ne 0^\circ}$)}
\begin{figure}
 \begin{center}
 \includegraphics[width=0.99\cwidth,trim=0 0 0 0,clip]{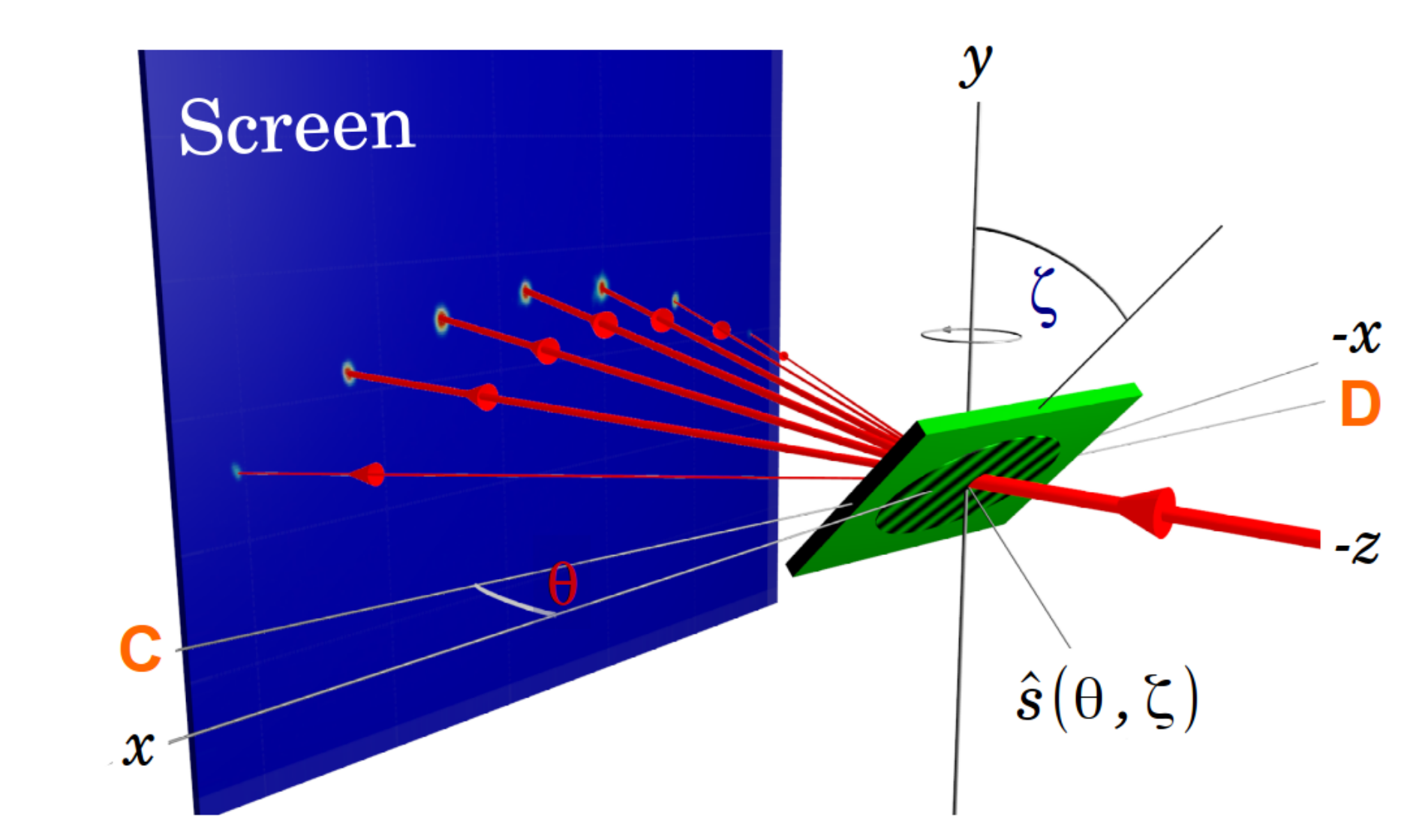}
\end{center}
 \caption{\label{fig:setup} Schematic of the setup with grating (green slab, center) and incident beam from the right (red). The outgoing, diffracted beams (red) proceed from the grating to the semitransparent screen (blue, left), where the positions of the diffraction spots form the diffraction pattern. Note that the coordinate system is defined as in Figs.\,\ref{fig:wavevec} and \ref{fig:coordsystem}.} 
 \end{figure}

\begin{figure}
\includegraphics[width=0.9\cwidth]{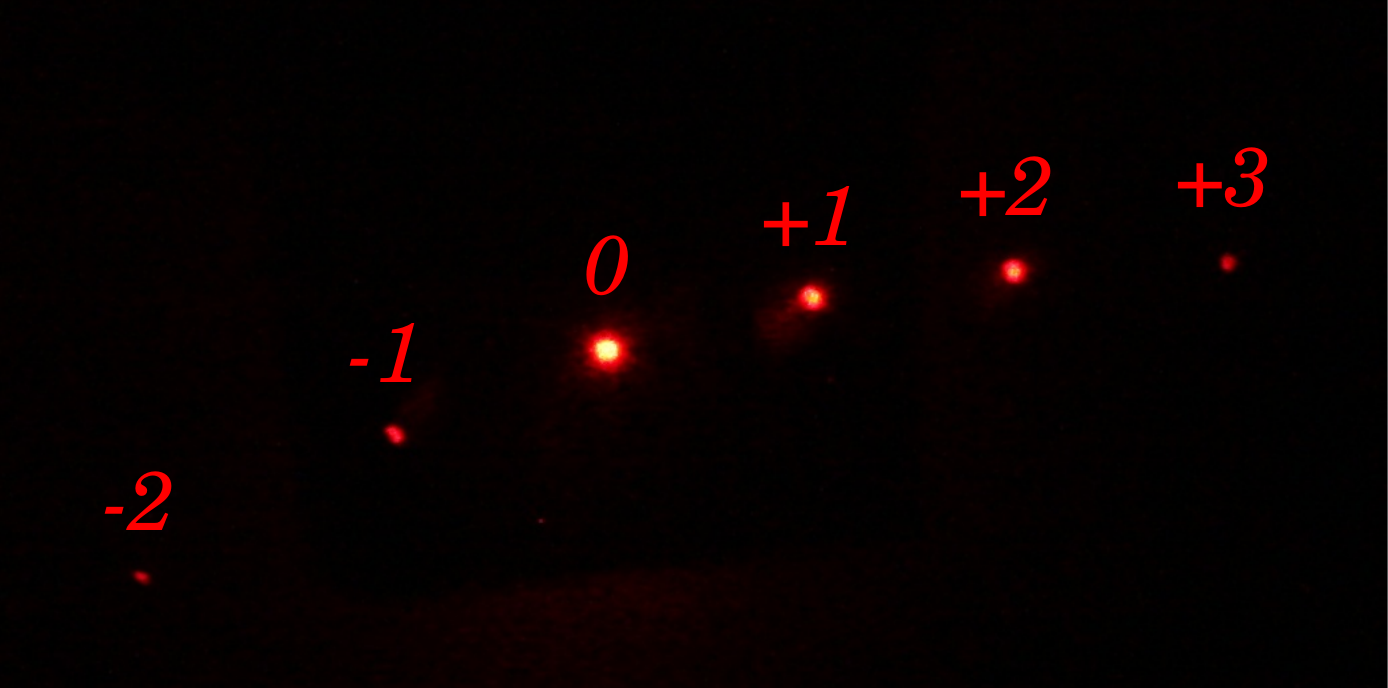}%
 \caption{\label{fig:art} Photograph of the diffraction pattern on the screen for  $\zeta=-43^\circ$ and $\theta=-20^\circ$. 
}
\end{figure}

The experimental setup to determine the directions of the diffracted beams
is shown in Fig.\,\ref{fig:setup}.
After setting the tilt around the grating vector $\vec G$ (axis {\sf CD}, fixed to the grating) by an angle $\zeta$, step-wise rotation about the $y$-axis to angles $\theta$ was performed. The distance between grating and screen was roughly 30\,cm. Photographs of the resulting diffraction patterns on the screen were taken for each value of $\zeta$ and $\theta$ to determine the position and the intensity of the spots. The experiments were carried out using a He-Ne laser (633 nm) at various tilting angles $\zeta$, in the range of rotation angles $\theta=-50^\circ\ldots+50^\circ$ with a step-width of $\Delta\theta=1^\circ$. In Fig.\,\ref{fig:art}, a photograph of the diffraction spots at $\theta=-20^\circ$ and $\zeta=-43^\circ$ is shown as an example.

It can be seen that -- at oblique incidence -- diffraction occurs out of the plane of incidence as soon as the Bragg condition is violated, as is expected from Eq.\,(\ref{eq:0}), which predicts off-plane diffraction for the order $m$ (with $m\ne 0$) when $\Delta k$ in Eq.\,(\ref{eq:mismatch}) is non-zero. 
The deviations of the diffracted beams' directions from the plane of incidence are different for the various diffraction orders with index $m$. For instance, in Fig.\,\ref{fig:art}, the beam positions corresponding to diffraction orders $+1,+2$ and $+3$, located right of the zero order beam position (the latter easily recognized here as the brightest spot), show little difference in their vertical off-plane coordinate (the $y$-coordinate). The contrary is the case for the beam positions corresponding to diffraction orders $-1,-2$, located left of the zero order beam position. For these positions, the differences in $x$ and $y$ coordinates are larger for different $m<0$ than they are for $m>0$.

\begin{figure}
\includegraphics[width=8cm]{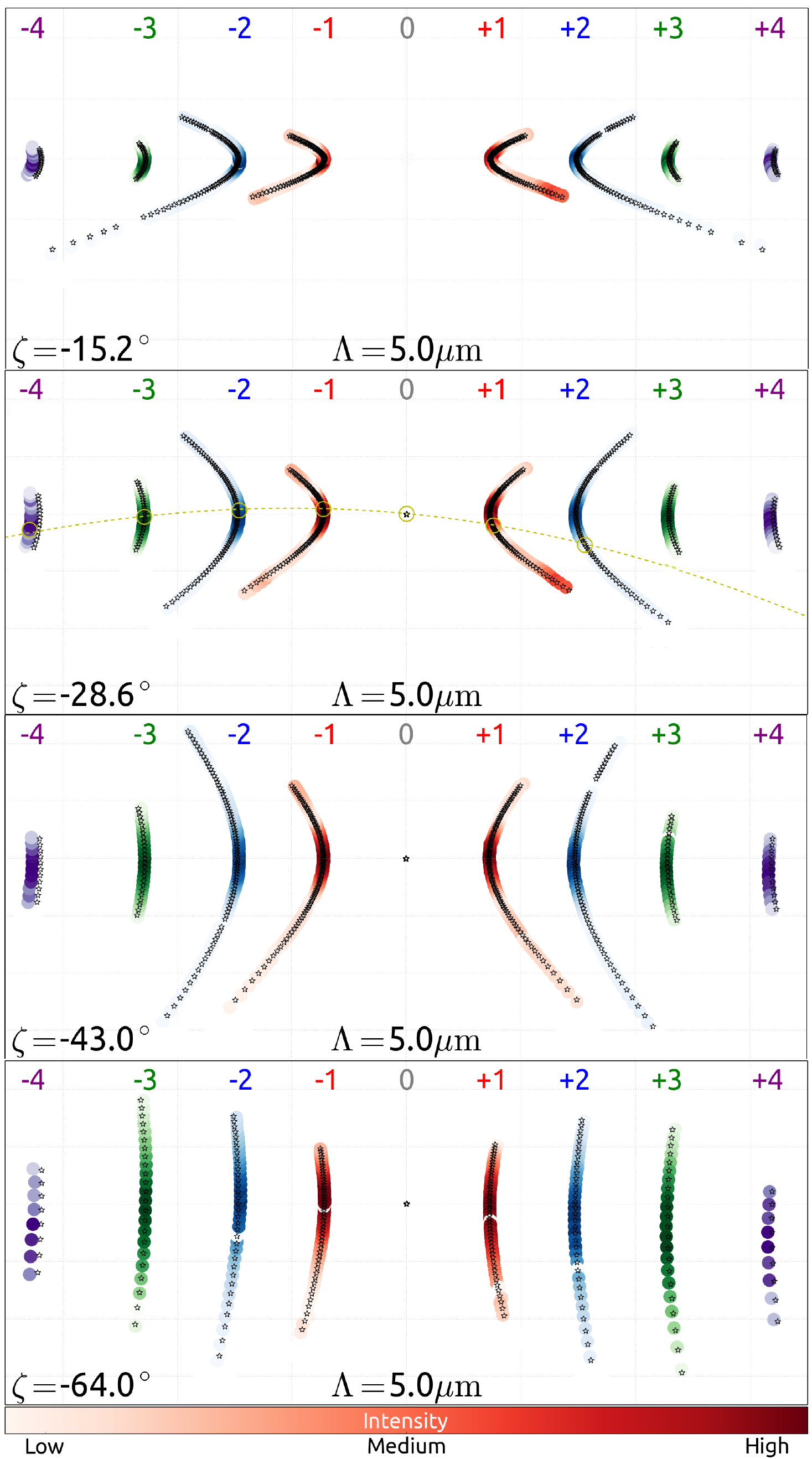}
\caption{\label{fig:posresults} Overlays of photographs of diffraction patterns (colored data points) on a screen (axes with arbitrary units) for $\theta$-scans (step-width $\Delta\theta=1^\circ$) at various tilt angles. The scale bar at the bottom is a reference for the intensity of the spots. Black, empty symbols (stars) are fits to the data according to Eqs.\,\ref{eq:positionsh} and \ref{eq:positionsv}\,(see text). 
The dashed line in the second overlay from top indicates the positions of intensity maxima in each diffraction order as calculated according to Eq.\,(17) of Ref.\,\cite{Jetty-ajp12}. Open circles correspond to our data at $\theta=-10^\circ$, for comparison (see text).} 
\end{figure}

In Fig.\,\ref{fig:posresults}, overlays of photographs of all diffraction patterns observed at each $\theta$ at tilt angles $\zeta=-15.2^\circ,-28.6^\circ-43.0^\circ
,-64.0^\circ$ are shown \footnote{The minus signs indicate that the tilt direction was chosen like in Fig.\,\ref{fig:setup}, rather than like in Figs.\,\,\ref{fig:coordsystem} and \ref{fig:wavevec}.}. One can clearly see that diffraction order maxima show up at a wide range of positions according to corresponding values of $\Delta k$ (and vectors $\hat s$) ranging from positive to negative while $\theta$ is variied at given $\zeta$.
As expected, the vertical off-plane component increases with increasing $\zeta$. Diffraction spots only up to the $\pm 4$th order could be observed, due to the very low diffraction signals at higher orders. The set of equations  
\begin{eqnarray}
h&=&L\frac{\Delta k(\theta,\zeta)\cos\zeta\sin\theta-m G \cos\theta}
{\beta+\Delta k(\theta,\zeta)\cos\zeta\cos\theta+m G\sin\theta}\label{eq:positionsh}\\
v&=&-L\frac{\Delta k(\theta,\zeta)\sin\zeta}{\beta+\Delta k(\theta,\zeta)\cos\zeta\cos\theta+m G\sin\theta}
\label{eq:positionsv}
\end{eqnarray} 
describing the horizontal and vertical positions of the diffraction spots was derived by use of Eqs.\,(\ref{eq:0}) and (\ref{eq:mismatch}). Here, $L$ is the distance from the grating to the screen. Equations\,(\ref{eq:positionsh}) and (\ref{eq:positionsv}) were fitted to the data, with $\zeta, \Lambda$ and $L$ set as free parameters. The parameter estimations are in good agreement with the measured values. The spot positions resulting from the fit are shown as black, empty symbols (stars) in Fig.\,\ref{fig:posresults}. 

Furthermore, in order to show in one example that also the Fresnel-Kirchoff-based approach agrees with our experimental data, we plotted the curve for the maximum-intensity spot positions according to  Jetty\,{\it et al.}\,\cite{Jetty-ajp12}, for $\zeta=-28.6^\circ$ and $\theta=-10^\circ$ as faint, dashed, yellow line in Fig.\,\ref{fig:posresults} (second from top). Open circles correspond to the measured spot positions for the $-4\ldots+2$ diffraction orders.

In Fig.\,\ref{fig:posresults} it can be seen that the larger $|m|$, the less diffraction maxima are observable at varying $\theta$: The far-out left and right `traces' of maxima in Fig.\,\ref{fig:posresults} contain very few points as compared to the low-$|m|$ orders.  
There are several possible reasons for this trend: One of them is total internal reflection by the glass-air interface at the back surface of the grating at large angles $\zeta$ and $\theta$. One less trivial reason for this behavior are non-propagating diffracted beams, as explained in the following:
Eq.\,(\ref{eq:mismatch}) predicts that for a given diffraction order index $m$, diffraction is observed only within certain angular limits at $\theta$ and $\zeta$, since -- when exceeding these limits -- $\Delta k$ and, thus, also the wavevector of the diffracted beam itself become complex. The corresponding $\vec k$-vectors describe non-propagating evanescent waves after the grating.
At given diffraction order $m$ and tilt angle $\zeta$, the analytic expression for the critcial angle $\theta_\text{\tiny{C}}$ is
\begin{eqnarray}
\sin\theta_\text{\tiny{C}}(\zeta,m)\!&=&\!-\frac{\rho_m}{\cos^2\zeta}\!+\!\frac{1}{2\cos^2\zeta}\nonumber\\
&\times&\!\!\sqrt{1\!+\!2\rho_m^2\!\!+\!\cos(2\zeta)\!\left[2\!+\!\cos(2\zeta)\!-\!2\rho_m^2
\right]}\label{eq:maxtheta}.
\end{eqnarray}

To answer the question on how many and which diffraction orders become propagating modes after the grating for a given geometry, one may set the square-root term in Eq.\,(\ref{eq:mismatch}) to zero. By solving for $m$, the limiting values $m_\pm$ are found, for which $\Delta k$ is just not yet complex. The maximum and minimum diffraction order indices $m_\pm$ at given $\theta$ and $\zeta$ can be written as
\begin{eqnarray*}
m_+\!&=&\!\left \lfloor{-\rho_2^{-1}\!\!\left(2\sin\theta\!+\!\sqrt{3\!+\!\cos^2\!\theta\cos(2\zeta)\!-\!\cos(2\theta)}\right)}\right \rfloor
\\
m_-\!&=&\!\left \lceil{-\rho_2^{-1}\!\!\left(2\sin\theta\!-\!\sqrt{3\!+\!\cos^2\!\theta\cos(2\zeta)\!-\!\cos(2\theta)}\right)}\right \rceil  
\end{eqnarray*}
where $\left \lceil{x}\right \rceil$ and $\left \lfloor{x}\right \rfloor$ denote the ceiling and floor functions, respectively.
The above equations for $m_\pm$ state that, at given 
$\theta$ and $\zeta$,
propagating waves (with real-valued wave vectors) corresponding to diffraction orders $m$ are excited, provided that their index $m$ lies between the two extremes, i.e., $m_{-}\leq m\leq m_+$. 

\section{Discussion}

We note that the behavior of diffraction patterns (positions of the diffraction spots) agrees well with Eqs.\,\ref{eq:positionsh} and \ref{eq:positionsv}, as can be seen in Fig.\,\ref{fig:posresults}. Our derivation of the relevant equations [Eqs.\,(\ref{eq:mismatch}) -- (\ref{eq:positionsv})] is based on energy- and momentum conservation only, using the Floquet theorem. However, solving Eq.\,(17) of the related, previous work by Jetty\,\emph{et al.}\,\cite{Jetty-ajp12} to obtain the vertical position of the spots on a screen as a function of the horizontal position, we find that the latter exactly matches the dependence of our Eqs.\,(\ref{eq:positionsh}) and (\ref{eq:positionsv}) in the limit of the slit height (see Ref.\,\cite{Jetty-ajp12}) approaching to zero. It is interesting and somewhat unexpected that, even if their work  \cite{Jetty-ajp12} and our experiment investigate off-plane diffraction in the context of very different diffraction regimes, the predictions and the data are similar and are in good agreement. Jetty\,\emph{et\,al.}'s experiment was clearly governed by the Raman-Nath regime in contrast to ours governed by the intermediate regime, where RCWA is necessary. 

It would be interesting to see if off-plane diffraction also occurs for massive particles, unlike photons. 
Considering experimental test of off-plane diffraction also for massive quantum objects like, for instance, neutrons, let us estimate the deviation from in-plane diffraction for the case of slow neutrons of de Broglie wavelength $\lambda_N=5$\,nm and grating spacing of $\Lambda=500$\,nm. 
The sample-detector distance $L$ is typically in the range of a couple of meters, say. The angular dependence curve of a grating with $d=30~\mu$m at $\zeta\approx 70^\circ$ shows an angular width (region around the diffraction maximum with acceptable intensity) of about $0.6^\circ$ (see, for instance, \cite{Fally-prl10}). Employing Eq.\,(\ref{eq:positionsv}), vertical shifts of the diffracted beams in the range of some 100 microns are expected, which can be detected with available neutron instrumentation and detector resolution. 

\section{Summary}
We have performed light optical diffraction experiments with a nanoparticle-polymer composite plane-wave grating. The angular dependence of the diffraction spots' positions at several angles of oblique incidence was fitted to the theoretical prediction derived from energy and momentum conservation and the proper boundary conditions, only. A comparison to a previous published study by Jetty\,\emph{et\,al.} \cite{Jetty-ajp12}, based on the Fresnel-Kirchhoff diffraction formula, yields perfect accordance in some special case. The latter is somewhat surprising, since we also demonstrate here that it is beyond the Fresnel-Kirchhoff approximation \cite{Born-02} or approximations such as the Raman-Nath transmittance theory \cite{Raman-piasa36,Raman.2-piasa36} to account for the angular dependence of the diffraction efficiency in a satisfying manner. In contrast, angular dependences of the diffraction efficiency calculated from measured intensities can be explained successfully using the RCWA \cite{Moharam-josaa95a}.

Finally, we have given an estimation for the size of the effect for neutrons, which suggests that a test of the phenomenon for matter waves is feasible with present-day technology.

\bibliographystyle{apsrev4-1}
%



\providecommand{\noopsort}[1]{}\providecommand{\singleletter}[1]{#1}%
\end{document}